\documentclass[12pt,letterpaper]{article}
\usepackage{graphics}
\usepackage{amssymb,epsfig,amsmath,euscript,array}
\usepackage{axodraw}
\usepackage{subfigure}
\usepackage{color}

\makeatletter  
\@addtoreset{equation}{section}
\makeatother

\newcounter{multieqs}

\def\bd{\begin{document}}
\def\ed{\end{document}}
\def\nn{\nonumber}
\def\bea{
\begin{eqnarray}}
	\def\eea{
\end{eqnarray}}
\let\bm=\bibitem
\let\la=\label

\begin{document}

\hfill{DESY 07-162}\\[-0.9cm]

\vspace{20pt}

\begin{center}

	{\huge\bf   Alpenglow\\--\\{\large A Signature for Chameleons in\\[1ex] 
                                        Axion-Like Particle Search Experiments}} \\[1.5ex] 

	\vspace{30pt}

	{M.~Ahlers, A.~Lindner, A.~Ringwald, L.~Schrempp, and C.~Weniger}

	{\small \em
	Deutsches Elektronen-Synchrotron DESY,\\
	Notkestrasse 85, D-22607  Hamburg, Germany}

	\vspace{40pt}

	{\bf Abstract}

\end{center}

We point out that chameleon field theories might reveal themselves as an ``afterglow'' effect 
in axion-like particle search experiments due to chameleon-photon conversion in a magnetic field. 
We estimate the parameter space which is accessible by currently available technology 
and find that afterglow experiments could constrain this parameter space in a way
complementary to gravitational and Casimir force experiments.
In addition, one could reach photon-chameleon couplings which are beyond the sensitivity of common laser 
polarization experiments. 
We also sketch the idea of a Fabry-P\'erot cavity with chameleons which could increase 
the experimental sensitivity significantly.

\setcounter{page}{0}
\thispagestyle{empty}
\newpage

\vspace{3ex}

\section{Introduction}

Both cosmology as well as popular extensions of the standard model of
particle physics are pointing to the possible existence of very weakly
interacting very light spin-zero (axion-like) particles (ALPs) and fields.
In fact, a plausible explanation for the apparent acceleration of the cosmic
expansion rate of the universe is provided by the presence of a spatially
homogeneous scalar field which is rolling down a very flat 
potential~\cite{Wetterich:1987fm,Ratra:1987rm,Caldwell:1997ii}. 
Remarkably, in string compactifications, there are many moduli fields
which couple to known matter with gravitational strength.    

Interactions of very light scalar fields with ordinary matter are strongly 
constrained by the non-observation of ``fifth force" effects 
leading to {\it e.g.}~violations of the equivalence principle. Correspondingly,
if such particles exist, the forces mediated by them should be either much weaker
than gravity or short-ranged in the laboratory. The latter occurs in theories 
where the mass of the scalar field depends effectively on the local 
density of matter -- in so-called chameleon field theories~\cite{Khoury:2003aq,Khoury:2003rn,Brax:2004qh}. 
Depending on the non-linear field self-interactions and on the interactions with the ambient 
matter, the chameleon may have a large mass in regions of high density (like the earth), 
while it has a small mass in regions of low density (like interstellar space).   
Since such kind of particle is able to hide so well from observations and experiments, 
it has been called a ``chameleon". 

Various phenomenological consequences of chameleon field theories have been discussed 
in the literature. Clearly, gravitational and other fifth force experiments are natural 
places where the effects of such particles can show up~\cite{Mota:2006ed,Mota:2006fz,Brax:2007vm}.
More recently, it has been emphasized~\cite{Jaeckel:2006xm,Brax:2007ak,Brax:2007hi} that chameleons may also 
be searched for 
in laser polarization experiments, such as BFRT~\cite{Cameron:1993mr}, BMV~\cite{Rizzo:2006bm}, 
PVLAS~\cite{Zavattini:2005tm, Zavattini:2007ee}, OSQAR~\cite{Pugnat:2006ba}, and Q\&A~\cite{Chen:2006cd}, 
which were originally planned for the laboratory search for axion-like particles based 
on the two photon couplings of the latter. 
In these experiments, polarized laser light is shone  
across a transverse magnetic field and changes in the polarization state are searched for.   
Similar to the case of standard ALPs, 
the real (virtual) production of chameleons would lead to an apparent
rotation (ellipticity) of the laser photons. 
But for chameleons the ellipticity can be much larger than the 
rotation~\cite{Brax:2007hi}.
This effect arises because chameleons are trapped inside the vacuum pipes~\cite{Jaeckel:2006xm}, due to their 
high effective mass in the walls.
Correspondingly, another type of laser experiments -- namely light-shining-through-a-wall experiments -- 
are not 
sensitive to chameleons. 
In these experiments, such as ALPS~\cite{Ehret:2007cm, ALPS:2007}, BMV~\cite{Rizzo:2006bm}, 
GammeV~\cite{GammeV:2007}, LIPSS~\cite{Baker:2006li}, OSQAR~\cite{Pugnat:2006ba} and 
PVLAS~\cite{Cantatore:2006pv}, laser light is shone onto a wall which separates the magnetic field into 
two regions and one searches for 
photons that might 
appear behind the wall due to ALP--photon conversion in the magnetic field. Clearly, chameleons do not reveal 
themselves in these kind of experiments, since they cannot pass through the wall.

\begin{figure}[p]
\begin{minipage}[c]{\linewidth}\centering
\begin{minipage}[c]{0.6\linewidth}\centering
\includegraphics[width=\linewidth]{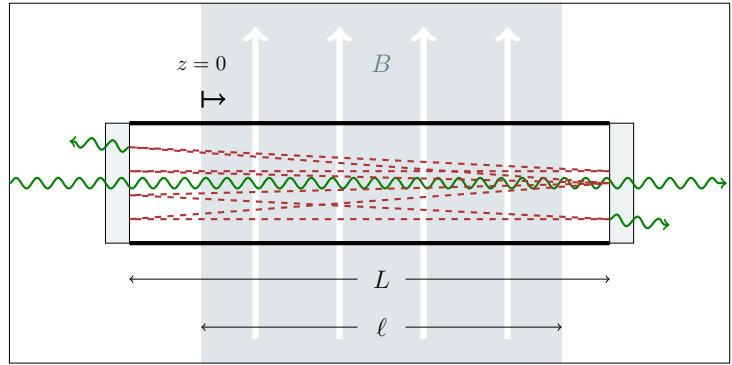}\\{\scriptsize (a)}
\end{minipage}\\[1cm]
\begin{minipage}[c]{0.6\linewidth}\centering
\includegraphics[width=\linewidth]{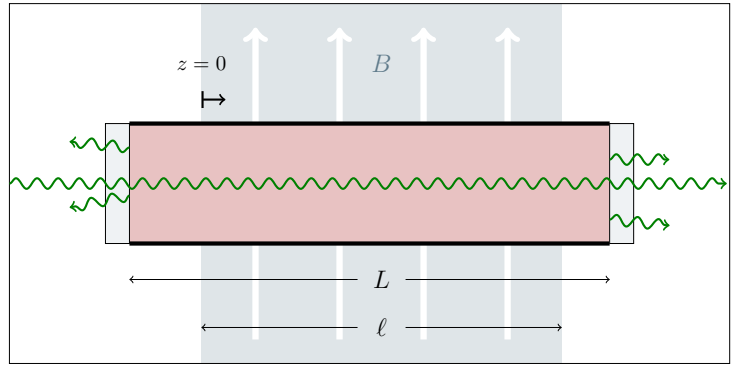}\\{\scriptsize (b)}
\end{minipage}\\[1cm]
\begin{minipage}[c]{0.6\linewidth}\centering
\includegraphics[width=\linewidth]{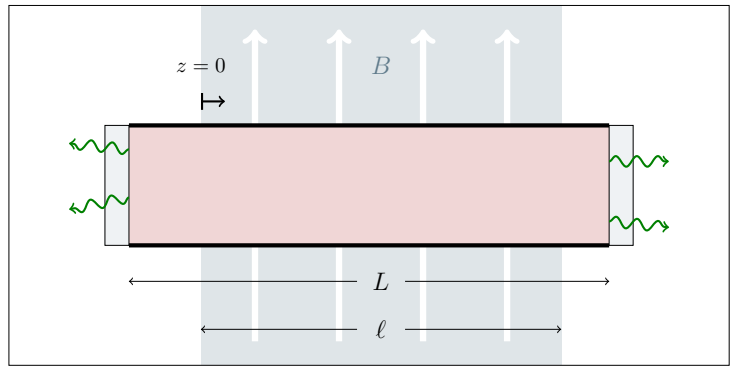}\\{\scriptsize (c)}
\end{minipage}
\end{minipage}
\caption[]{\small Illustration of an afterglow experiment to search for chameleon particles.
(a) Filling the vacuum tube by means of a laser beam with chameleons 
via photon-chameleon conversion in a magnetic field.
(b) An isotropic chameleon gas forms.
(c) Afterglow from chameleon-photon conversion in a magnetic field.}
\label{fig:Setup}
\end{figure}

In this letter, we want to propose the search for a unique signature of chameleons in the above mentioned 
axion-like particle search experiments: an ``afterglow'' effect\footnote{A similar proposal 
can be found in Ref.~\cite{Brax:2007}.}.

The basic idea is depicted in Fig.~ \ref{fig:Setup}.
In the beginning (a), a laser beam is shone through a vacuum cavity in a strong magnetic field. The end 
caps 
of the cavity are transparent to laser photons. Some of the photons are converted to chameleon particles 
whose mass $m$ in the vacuum is smaller than the energy $\omega$ of the laser beam. However, both the matter 
density and thus also $m$ are much higher within the end caps such that there $m\gg\omega$ holds. As a 
consequence, the chameleons 
are then reflected by these end caps and remain trapped inside the 
cavity~\cite{Jaeckel:2006xm,Brax:2007hi}, which gets more and more 
filled (b), whereas the photons escape. 
After switching off the laser (c), one can search for afterglow photons, that emerge from the 
back-conversion of the trapped chameleon particles in the magnetic field.

The organization of the paper is as follows. Firstly, we introduce the basic idea and some properties of 
chameleon theories and derive the conversion probability between photons and chameleons from their 
equations of motion. Secondly, we consider in some detail the loading and unloading rate
of vacuum cavities and estimate the parameter space, which is accessible by afterglow experiments, using 
the experimental parameters of the ALPS~\cite{Ehret:2007cm, ALPS:2007} experiment. 
We also discuss the possibility of a coherent evolution of the chameleon wave, which would drastically 
increase the sensitivity of the experiment. Finally, we state our 
conclusions and give an 
outlook on open questions.

\section{Chameleon Theory}

Chameleon theories can be described as scalar field theories with a self-interaction potential $V(\phi)$ 
for the chameleon $\phi$ and a conformal coupling to matter. More precisely, the dynamics of a chameleon 
field in the presence of matter and electromagnetism are governed by the general action,
\begin{eqnarray}
	\label{action}
S=\int d^4x \sqrt{-g}\left(-\frac{M^2_{\rm Pl}}{16\pi} R
+\frac{1}{2}(\partial \phi)^2 -V(\phi)
-\frac{e^{2\phi/M}}{4} F^2\right)
+ \sum S_{\rm m}^{(i)}( e^{2\phi/M_i}
g_{\mu\nu},\psi_{\rm m}^{(i)}), \nonumber\\
\end{eqnarray} 
where $M_{\rm Pl}=1/\sqrt{G}\simeq 1.2\cdot 10^{19}$ GeV is the Planck mass, $g$ is the determinant of 
the metric $g_{\mu\nu}$, ${ R}$ is the Ricci-Scalar, and $F_{\mu\nu}$ denotes the electromagnetic 
field strength tensor. Furthermore, the chameleon couples conformally both to electromagnetism and 
to the various matter fields $\psi_{\rm m}^{(i)}$ through the dependence of their matter actions 
$S_m^{(i)}$ on $e^{2\phi/M_i}$. For simplicity, in the following, we will assume the chameleon to 
exhibit a universal coupling to matter and to electromagnetism which implies $M_i=M$. 
As we will see later on, in order to expect detectable signatures of a chameleon theory in 
axion-like particle search experiments, one has to require 
$M_{\rm Pl}\gg M$. 

As can be deduced from varying the action~\cite{Khoury:2003rn}, the dynamics of the chameleon field are governed 
by an effective potential $V_{\rm eff}(\phi)$. In the presence of (non-relativistic) matter and external 
electric ($\vec{E}$) and magnetic ($\vec{B}$) fields, it takes the following form,
\begin{equation}\label{Veff}
V_{\rm eff}(\phi)=
V(\phi) + \rho_{\rm eff}\,e^{\phi/M}
, 
\end{equation}
where
\begin{equation}\label{Vrhoeff}
\rho_{\rm eff} 
=\rho_{\rm matter}-\frac{1}{2}\left( \vec{E}^2 -\vec{B}^2\right)
.
\end{equation} 
In order to ensure that $\phi$ exhibits the 
characteristic environmental dependencies of a chameleon field, we require its self-interaction 
potential to be of a runaway form in the sense that
\begin{equation}\label{condV}
V'(\phi)<0,\,\,V''(\phi)>0,\,\,V'''(\phi)<0,
\end{equation}
with $V'=dV/d\phi$. Accordingly, in low matter density environments (like in outer space) the dynamics of 
$\phi$ are determined by its quintessence-like self-interaction potential. However, even though $V(\phi)$ 
is monotonically decreasing, for our exponential choice of the conformal coupling (cf. Eq.~\eqref{action}) in 
over-dense regions the chameleon is stabilized in a minimum exerted by its effective potential $V_{\rm eff}$. 
More precisely, according to Eqs.~\eqref{Veff} and \eqref{condV}, the local effective density 
$\rho_{\rm eff}$ governs both 
the location of the effective minimum at $\phi=\phi_{\rm min}$ where $V'_{\rm eff}(\phi_{\rm min})=0$ and 
the mass $m^2=V''(\phi_{\min})$ of the chameleon field, which grows with increasing matter density. 
To see this explicitly, for definiteness, we take a self-interaction potential of the form,
\begin{equation}\label{V}
V(\phi)=\Lambda^4e^{\Lambda^n/\phi^n}\approx \Lambda^4+\frac{\Lambda^{n+4}}{\phi^n},
\end{equation}
where in the last equality and throughout this work we assume $\phi\gg \Lambda$ such that $V$ reduces 
to a constant term plus a Ratra-Peebles inverse power law potential. This form of the potential has been 
shown not to be in conflict with current laboratory, solar system or cosmological 
experiments~\cite{Khoury:2003aq,Brax:2004qh} even in the strong coupling case, 
$M_{\rm Pl}\gg M$~\cite{Mota:2006fz}. Note that in order for the chameleon to drive cosmic acceleration,
one has to require $\Lambda\approx 2.3$~meV such that the $\Lambda^4$ term in Eq.~\eqref{V} acts as an 
effective cosmological constant.
 
Since limits on any variation of fundamental constants of nature 
require $\phi/M\ll 1$~\cite{Mota:2006fz}, we henceforth take $\exp(\phi/M)\approx 1+\phi/M$ 
for the conformal matter coupling. Under these assumptions, Eq.~\eqref{Veff} yields 
\begin{equation}
V'_{\rm eff}(\phi_{\rm min})=0=V'(\phi_{\rm min})+\frac{\rho_{\rm eff}}{M},
\end{equation}
which defines
\begin{equation}
\phi_{\rm min}=\left(\frac{n M \Lambda^{n+4}}{\rho_{\rm eff}}\right)^{1/(1+n)}.
\end{equation}
Thus, the mass of the scalar field follows from,
\begin{equation}\label{mphi}
m^2=\left(\frac{n(n+1)\Lambda^{n+4}}{\phi_{\rm min}^{n+2}}\right)=
n(n+1)\Lambda^2\left(\frac{\rho_{\rm eff}}{n M \Lambda^3}\right)^{(n+2)/(n+1)}.
\end{equation}
Accordingly, we note that the chameleon mass in an axion-like particle search experiment depends on the 
details of the experimental set-up. Namely, it is determined by the effective density $\rho_{\rm eff}$ . 
Furthermore, it depends on the size of the cavity in the following sense. If the radius of the cavity 
$r$ is too small, $r\lesssim 2/m(\phi_{\rm min})$, the chameleon field will not 
relax to $\phi_{\rm min}$ and thus its mass turns out to only depend on the geometry of the 
set-up~\cite{Brax:2007hi}, bounded from below like 
\begin{equation}
\label{eqn:geomass}
m\gtrsim\frac{2\sqrt{n+1}}{r}=2.3 \cdot 10^{-5}\,{\rm eV}\,\sqrt{n+1}\left(\frac{1.7\,{\rm cm}}{r}\right).
\end{equation}

\section{Afterglow Experiment}\label{sec:afterglow}

The chameleon--photon interaction in a magnetic field can be described as a two-level systems as in 
Ref.~\cite{Raffelt:1987im} (see also Refs.~\cite{Brax:2007hi,Brax:2007}). For the 
discussion of coherent chameleon interactions with the laser following in the next section it is 
convenient to repeat the basic steps of the calculation.

We consider a linearly polarized laser beam propagating perpendicular to a magnetic field $\vec{B}$ 
oriented orthogonal to the polarization vector of the laser. The equations of motion for right-moving 
($-$) and left-moving ($+$) plane waves $(A,\phi)^T$ are given as
\begin{equation}
\left[\omega\mp i\partial_z -\begin{pmatrix}0& 
\mp\Delta_M\\\mp\Delta_M&\Delta_\phi\end{pmatrix}\right]\begin{pmatrix}A\\\phi\end{pmatrix}=0
\end{equation}
with the inverse oscillations lengths $\Delta_\phi=m^2/2\omega$ and $\Delta_M=B/2M$. 
These equations can be diagonalized by a rotation 
\begin{equation}
\begin{pmatrix}A'\\\phi'\end{pmatrix}=
\begin{pmatrix}\cos\varphi&\pm\sin\varphi\\\mp\sin\varphi&\cos\varphi\end{pmatrix}
\begin{pmatrix}A\\\phi\end{pmatrix}\equiv\mathcal{Q}(\pm\varphi)\begin{pmatrix}A\\\phi\end{pmatrix}
\end{equation}
with $\frac{1}{2}\tan2\varphi = \Delta_M/\Delta_\phi$. The rotated fields have the solution
\begin{equation}
\begin{pmatrix}A'(z)\\\phi'(z)\end{pmatrix}=\begin{pmatrix} e^{\mp i k_-z}&0\\0&e^{\mp i k_+z}\end{pmatrix}
\begin{pmatrix}A'(0)\\\phi'(0)\end{pmatrix}\equiv\mathcal{M}'(\pm z)\begin{pmatrix}A'(0)\\\phi'(0)\end{pmatrix}
\end{equation}
with $k_\pm = \omega-\frac{\Delta_\phi}{2}(1\pm\sec2\varphi)$. Note that for small mixing $\varphi$ the 
state $A'$ is photon-like with $k_-\approx\omega$ whereas $\phi'$ propagates with 
$k_+\approx\omega-\Delta_\phi$. Hence, in the small mixing case, the coherence length of the 
chameleon-photon system will be $\ell_\text{coh}=\pi/\Delta_\phi$. The transition probability 
$P_{\gamma\to\phi}(\ell,B)=P_{\phi\to\gamma}(\ell,B)=1-P_{\phi\to\phi}(\ell,B)$ is given by
\begin{equation}\label{prob}
P_{\gamma\to\phi}(\ell,B) = |(\mathcal{Q}^T(\varphi)\mathcal{M}'(\ell,B)\mathcal{Q}(\varphi))_{21}|^2 
= 4 \left( \frac{B\omega}{Mm^2} \right)^2 \sin^2 \left(\frac{\ell m^2}{4\omega}\right) 
+ \mathcal{O}\left(\varphi^4\right)\,\,.
\end{equation}
As noted in Ref.~\cite{Jaeckel:2006xm,Brax:2007hi}, the walls of a vacuum tube can act as mirrors 
for chameleon particles if their energy is smaller than their effective mass in the cavity wall. Since this is 
the case in common laser experiments with vacuum tubes, chameleon particles are trapped in the cavity 
once they are produced and can only escape via back-transition to photons. In the following we will treat 
the chameleons trapped inside the tube as a ``gas'', neglecting interference effects with the incoming 
laser field. 

\subsection{Chameleon Generation}

As a first step, consider a set-up (similar to the one in Fig.~\ref{fig:Setup}) where a linearly polarized 
laser beam with power $\mathcal{P}_\gamma$ and frequency $\omega$ enters a vacuum between two 
(infinitely extended) windows separated by a distance $\ell$. We assume that a magnetic field $B$ 
is oriented parallel to the window surfaces and
orthogonal to the polarization vector of the laser beam. For simplicity, we also assume that the 
magnetic field region is confined to the vacuum between the windows. The laser beam enters through 
one of the windows with an angle $\theta$ with respect to the window surfaces.

After switching on the laser beam, the number of laser photons between the windows is $\ell' I_\gamma$, 
where
the photon current $I_\gamma$ is defined as $I_\gamma=\mathcal{P}_\gamma / \omega$, and 
$\ell'=\ell/ \sin\theta$ is the length of the laser beam between the windows.
A fraction of $P_{\gamma\rightarrow\phi}(\ell', B')$ laser photons transforms to chameleon particles 
where $B'=B\sin\theta$.
These chameleon particles are then reflected back and forth between the windows and become trapped whereas the 
photons escape.  At each reflection, a fraction of 
$P_{\phi\rightarrow\gamma}(\ell', B')$ chameleons are lost due to back transition to photons. 
Taking $N$ reflections into account, the total number of chameleons stored in the cavity is then given by
\begin{equation}
N_\phi = \ell' P_{\gamma\to\phi}\sum_{k=0}^{N-1}P^k_{\phi\to\phi} I_\gamma= \ell' 
P_{\gamma\to\phi}\left(\frac{1-P^{N}_{\phi\to\phi} }{1-P_{\phi\to\phi}}\right) I_\gamma\,,
\end{equation}
where $P_{\phi\to\phi}=1-P_{\phi\to\gamma}$ and we approximate the velocity of the chameleons by
the speed of light. In the limit of large $N$, it follows that
\begin{equation}
N_\phi =  \ell' \left(1-P^{N}_{\phi\to\phi}\right)I_\gamma\xrightarrow{N\to\infty} \ell' I_\gamma\,,
\end{equation}
{\it i.e.} the total number of chameleon particles equals the number of photons inside the cavity, $\ell' I_\gamma$.

Identifying $t=\ell'N$, the rate $\Gamma$ to load and unload the cavity with chameleons is given by
\begin{align}\label{Tdecoh}
	\Gamma = -\frac{\ln P_{\phi\to\phi}}{\ell'}=-\frac{\ln\left(1- P_{\phi\to\gamma}\right)}{\ell'}
\simeq\frac{ P_{\phi\to\gamma}}{\ell'}\simeq\frac{4}{\ell'}\left(\frac{\omega B'}{Mm^2}\right)^2
\sin^2\left(\frac{\ell' m^2}{4\omega}\right)\,,
\end{align}
which results in an (un-)loading time
\begin{equation}
\tau_{\rm load}\equiv\frac{1}{\Gamma} 
\simeq \frac{4M^2}{B'^2\ell'}\simeq 111\ {\rm s}\left( \frac{M}{10^6\,{\rm GeV}}\right)^2
\left( \frac{5.4\,{\rm T}}{B'}\right)^2 \left( \frac{4.2\,{\rm m}}{\ell'}\right) , 
\end{equation}
for $m\lesssim 2\sqrt{\omega/\ell^\prime}=6.6\cdot 10^{-4}\,{\rm eV}\,\sqrt{\omega/2.3\,{\rm eV}}
\,\sqrt{4.2\,{\rm m}/\ell'}$. 
For later convenience, we define the {\it characteristic length scale} $R$ of the cavity as 
\begin{equation}
	R\equiv\Gamma \cdot \frac{4M^2}{B^2}.
	\label{eqn:DefR}
\end{equation}
For small masses $m$, it is given by $R\simeq \ell \sin\theta$ in the above example. For large masses, 
$R$ strongly oscillates between zero and $R=4(B')^2\omega^2/M^2m^4\ell'$ as a function of $\ell$.

To determine $R$ for a more realistic set-up, note that chameleons become reflected by the cavity walls 
approximately $10^8(\ell/1\,\text{m})^{-1}$ times per second. If the walls of the cavity are not aligned 
to be exactly 
parallel to each other (with angular deviations of the order of $\Delta\delta$), trapped particles will 
become distributed isotropically after some time $t\simeq \mathcal{O}(\ell/\Delta\delta)$.  
In realistic cases, 
especially if no extra effort is put into the adjustment of the cavity walls, $t$ is well below 
$\sim\mathcal{O}(1\,\text{s})$.
Since the characteristic time scale of an afterglow experiment will turn out to be orders of magnitude 
larger, it seems to be safe to assume that the chameleons form an isotropic and homogeneous gas inside 
the cavity.

In such a set-up, the number of chameleons in the cavity is governed by a simple rate equation:
\begin{equation}
	\frac{dN_\phi}{dt}=P_{\gamma\rightarrow\phi}(\ell, B) I_\gamma - \Gamma N_\phi.
	\label{eqn:RateEquation}
\end{equation}
The first term on the right-hand side describes the loading of the cavity, where $\ell$ now is the length 
of the laser beam inside the magnetic field in the cavity (see Fig.~\ref{fig:Setup}). 
The second term describes the loss of chameleons due to back-conversion to photons. 
The corresponding (un-)loading rate $\Gamma$ is given by
\begin{equation}
        \Gamma=\frac{1}{4\pi V}\int dA \int_{2\pi} d\Omega \cos\theta_S P_
{\phi\rightarrow\gamma}(\ell'(x,e_\Omega), B\cdot \sin\theta ),
        \label{eqn:GammaDef}
\end{equation}
where $V$ is the volume of the cavity, $\int dA$ denotes the integration
over the cavity surface, $\int d\Omega$ the
integration over the half sphere in which photons can be emitted from
each point of the surface, $\ell'(x,e_\Omega)$
is the distance a particle at surface point $x$ and orientation $e_
\Omega$ has traveled inside the magnetic field since its last
reflection, $\theta$ denotes the angle between $e_\Omega$ and the
direction of the magnetic field and $\theta_S$ denotes the angle between
$e_\Omega$ and the surface normal. 

In this definition we have assumed that the cavity is completely transparent to photons, and that 
the chameleon-photon wave is projected onto its chameleon (or photon) part at each reflection.
Furthermore, we have neglected all interactions between chameleon particles and assumed that they can propagate
freely in the cavity without thermalization\footnote{Since the expansion of the effective potential 
$V_\text{eff}(\phi)$ (see Eq.~\eqref{Veff}) around its minimum contains interaction
terms of all powers of $\phi$, processes like $\phi\phi\to n\, \phi$ are allowed for all $n$ even 
at tree level. 
These chameleon creation and annihilation cross sections might become very large for large 
energies and thus 
lead to thermalization or
to a breakdown of the particle description due to strong coupling for certain values of $M$ and $\Lambda$. 
In this letter we only consider the idealized case without thermalization 
and postpone the further investigation of these problems to a forthcoming publication~\cite{We:2007}.}. 
Since the energy of chameleon particles which are produced by laser beams corresponds to temperatures 
of the order of $T\sim\mathcal{O}(10^4\text{K})$, one might fear that they quickly lose their energy 
to the cavity walls and take on the temperature of the surrounding. However, this process seems to be 
negligible\footnote{
	Chameleons in our experiment are expected to be much hotter than the cavity walls since the 
laser frequency $\omega=2.3\,\text{eV}$ corresponds to a temperature of $2.7\cdot 10^4\, \text{K}$.
	To estimate the energy loss of the chameleons through elastic scattering off the cavity walls, 
note that the maximal energy transfer from a chameleon particle with energy $\omega$ on a part of the 
wall with mass $M$ is $\Delta\omega \simeq 2\omega^2/M$. 
	$M$ may be approximated by the mass of a half sphere with diameter of the wavelength of the 
chameleon $\lambda=532\,\text{nm}$ and depends on the mass density of the wall, $\rho_\text{wall}$. 
	From that we obtain a cooling time of
	$
		t_\text{cooling}= 2.4\cdot 10^{8}\,\text{s} 
\left( \rho_\text{wall}/1\text{g}\, \text{cm}^{-3} \right)\left( R/1\text{cm} \right),
		\label{}
	$
	where $R$ denotes the average path length of a chameleon between two reflections. 
Since $R\gtrsim 1\,\text{cm}$ and $\rho_\text{wall}\gtrsim 1\,\text{g\,cm}^{-3}$, we conclude that this effect 
is negligible.}.
Finally, the description of chameleons as particles with constant mass certainly breaks down for parameters
$(M,\Lambda)$ for which Eq.~\eqref{mphi} yields masses below $\sim 2/r$. In these cases, the real chameleon mass
strongly depends on the position in the cavity and is constrained from below as 
in Eq.~\eqref{eqn:geomass}. Since large parts of the corresponding parameter space are already excluded 
by other experiments (at least for $n=1$, see region above the ALPS line Fig. \ref{fig:Bounds}), and for 
simplicity, we exclude this region from our considerations.

For the case of a tube with length $L$, radius $r\ll L$ and a magnetic field of length $\ell$ 
(see Fig.~\ref{fig:Setup}), one obtains
\begin{equation}
\Gamma= \frac{B^2}{4M^2} \cdot\begin{cases}\frac{2r\ell}{L}
\left( 1-\frac{r}{2\ell}\ln \frac{\ell}{r} \right),&
\quad\text{for}\hspace{1ex}m^2\lesssim 4\omega/\ell\,,\\
\frac{2r\ell}{L}
\left(1-\frac{rm^2}{8\omega}\ln \frac{4\omega}{rm^2} \right),
&\quad\text{for}\hspace{1ex}4
\omega/\ell\lesssim m^2\,\ll\, 4\omega/r\,,\\
\frac{5}{2}\frac{\ell\omega^2}{Lrm^4},&\quad\text{for}\hspace{1ex}m^2\,\gg\, 4\omega/r\,.
\end{cases}
\label{eqn:GammaLog}
\end{equation}
From the previous equation and Eq.~\eqref{eqn:DefR} the characteristic length scale $R$ of the system 
can be directly read off. Note, that if $m^2\lesssim 4\omega/\ell$, $\Gamma$ is independent of 
the chameleon mass, while, for $m^2 \gg \omega/r$, $\Gamma$ decreases like $\sim m^{-4}$.

As we stated above, the formulae for $\Gamma$ only hold, if the chameleon-photon wave is projected onto 
its chameleon or photon part at each reflection. This is certainly the case if the walls are transparent 
to photons, since the wave packets of the 
combined chameleon-photon system become separated after the reflection. Due to interaction with the 
environment, this spatial separation acts like a projection onto either the photon or chameleon part.
However, this measurement-like process might even occur if photons {\it are} reflected by the cavity 
walls ({\it e.g.}~if the cavity is a metallic tube where only the end caps are transparent to photons). 
This is due to the fact, that the reflection properties of chameleons and photons are in general 
quite different \cite{Brax:2007hi}, so that the corresponding wave-packets may become separated. 
This separation again acts like a measurement and leads to a projection onto either the photon or the 
chameleon part.

\subsection{Photon Regeneration}

We now turn the determination of the results which can be expected in an afterglow experiment. 
Switching on the laser for a time $\Delta t$ fills the cavity with 
\begin{equation}\label{eqn:Ntot}
	N_\phi^0(\Delta t)= 
 \frac{\left( 1-e^{-\Gamma \Delta t} \right)}{\Gamma}\,P_{\gamma\rightarrow\phi}(\ell) I_\gamma 
\end{equation}
chameleon particles. Assuming, that a fraction of $f$ regenerated photons hit the detector, the number of 
detected photons per second
immediately after switching off the laser (at time $t=0$) is
\begin{eqnarray}
	I_\gamma^{\text{det}}(t=0) = 
f  
P_{\gamma\rightarrow\phi}(\ell) 
 I_\gamma\, \left( 1-e^{-\Gamma \Delta t} \right) 
= f P_{\gamma\rightarrow\phi}(\ell)I_\gamma \cdot\begin{cases}\Gamma \Delta t, 
&\quad\text{for}\hspace{1ex}\Gamma \Delta t\lesssim 1\,,\\
1,&\quad\text{for}\hspace{1ex}\Gamma \Delta t\gtrsim 1\,.
\end{cases}
	\label{eqn:Idet}
\end{eqnarray}
The current of detected photons decreases with measurement time like 
$e^{-\Gamma t}$. 
Hence, the number of detected photons for a measurement time $t$ is 
\begin{equation}
	N_\gamma^\text{det}(t)=f\, 
P_{\gamma\rightarrow\phi}(\ell) I_\gamma\,  \left( 1-e^{-\Gamma \Delta t} \right)\,
\frac{\left( 1-e^{-\Gamma t} \right)}{\Gamma}.
\end{equation} 
If the detector is sensitive down to $N_\gamma^\text{sen}$, the experiment can in 
principle reach parameters $(M, \Lambda)$ for which $N_\gamma^\text{det}(t)\gtrsim N_\gamma^\text{sen}$. 
This region is
further constrained by the requirement, that a chameleon inside the cavity cannot have a mass smaller than 
$m\sim 2/r$ (see above). 

\begin{figure}[p]
\begin{minipage}[c]{\linewidth}\centering
\includegraphics[width=0.95\linewidth]{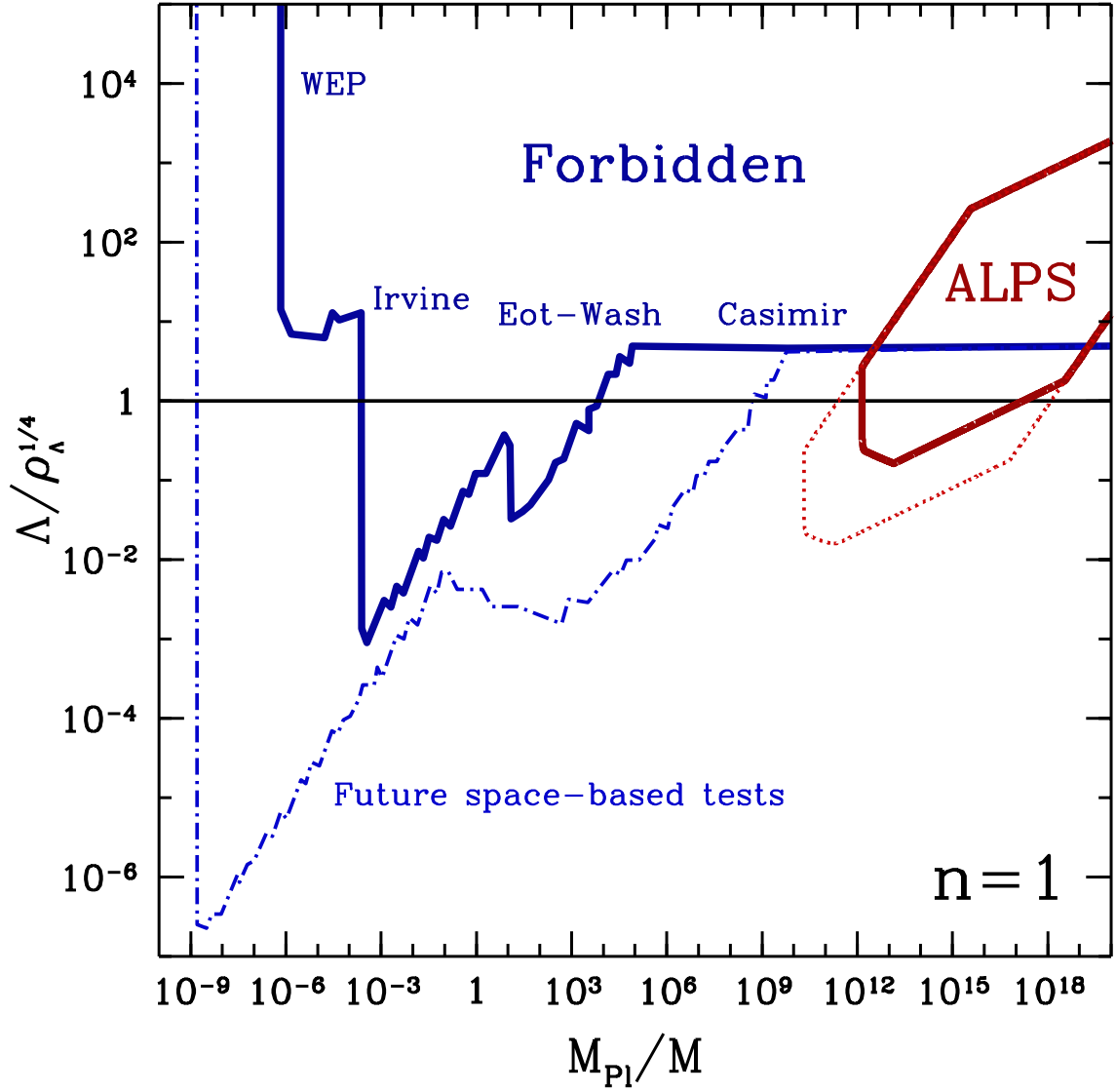}
\end{minipage}
\caption[]{\small
Combined constraints on $n=1$ chameleon theories in terms of the potential parameters $\Lambda$, 
in units of the energy scale $\rho_\Lambda^{1/4}\simeq 0.002\ {\rm eV} \simeq (0.1\ {\rm mm})^{-1}$ 
corresponding to the energy density of dark energy, versus the inverse coupling scale $1/M$, in 
units of  $M_{\rm Pl}\simeq 1.2\cdot 10^{19}$\,GeV. 
Current constraints (blue solid line; from Ref.~\cite{Mota:2006fz}) arise from 
searches for violations of the weak equivalence principle (WEP), from searches for deviations from the 
$1/r^2$ law of the gravitational force (Irvine and E\"ot-Wash), and from bounds on the strength of any 
fifth force from measurements of the Casimir force. Future space-based tests will improve the current 
limits to the one shown as a light-blue dotted dashed line. A search for an afterglow due to chameleon-photon 
reconversion at ALPS phase zero will be sensitive to the region indicated by the red solid
line, corresponding to a loading and measurement time of $\Delta t = t = 100$\,s.     
The red dotted line corresponds to the academic case $\Delta t = 1$\,y and $t=1$\,h.}
\label{fig:Bounds}
\end{figure}

We now turn to an estimate of the coupling parameter $M$ 
accessible with a set-up similar to the ALPS phase zero 
experiment~\cite{ALPS:2007}. Using the corresponding experimental parameters ($\omega=2.3\,\text{eV}$, 
$\ell=4.2\,\text{m}$, $L=6.5\,\text{m}$, $r=1.7\,\text{cm}$, and $B=5.4\,\text{T}$),  
we can determine the (un-)loading 
rate $\Gamma$ from Eq.~\eqref{eqn:GammaLog} and obtain
\begin{equation}
	\Gamma\simeq 4.6\cdot 10^{-5}\,\text{s}^{-1} \left(\frac{10^6\,\text{GeV}}{M}  \right)^2,
	\hspace{1cm}\text{for}\hspace{1ex}
	m\lesssim 2\sqrt{\omega/\ell}=6.6\cdot 10^{-4}\,\text{eV}.
	\label{eqn:GammaWert}
\end{equation}
In the same mass limit, the conversion probability that enters Eq.~\eqref{eqn:Idet} is given by
\begin{equation}
	P_{\gamma\to\phi}(\ell, B)= 1.3\cdot 10^{-10} \left(\frac{10^6\,\text{GeV}}{M} \right)^2.
	\label{}
\end{equation}
In ALPS phase zero the laser has a power $\mathcal{P}_\gamma=3\,\text{W}$,  
corresponding to a photon current of $I_\gamma\simeq 8.1\cdot 10^{18}\,\text{s}^{-1}$. 
The exploited CCD camera is sensitive down to $N_\gamma^\text{sen}=5\cdot 10^5$.   
For small masses, an infinite loading time $\Delta t$, and $f\simeq 1/2$, 
the experiment could reach in principle coupling parameters $M$ 
up to $M\simeq 10^7\,\text{GeV}$. This value scales with the square root of the laser power, 
stronger lasers
easily reach $M\gtrsim 10^8\,\text{GeV}$. 
However, in these cases the loading time becomes very large, since it scales with $\sim M^{2}$ 
(cf.~Eq.~\eqref{eqn:GammaWert}). 
Hence, the sensitivity of afterglow experiments is mainly constrained by the finite loading  
time of an experiment.

Let us now consider also the sensitivity to the parameter $\Lambda$ in the self-interaction
potential~\eqref{V} of the chameleon which enters here through the chameleon mass~\eqref{mphi}.
The residual gas density in the vacuum tube of the ALPS experiment is 
$\rho_\text{gas}\simeq 2\cdot 10^{-14}\,\text{g\,cm}^{-3}$.
Together with the magnetic field this yields an effective energy density of 
$\rho\simeq 1.3\cdot 10^{-13}\,\text{g\,cm}^{-3}$. 
From this we can determine the parameter region, which is accessible with the ALPS phase zero set-up.
In Figure~\ref{fig:Bounds}, we have displayed two cases: the red solid
line corresponds to a loading and measurement time of $\Delta t = t = 100$\,s, while the     
red dotted line shows the academic case $\Delta t = 1$\,y and $t=1$\,h. 
Evidently, already in the ALPS phase zero set-up which is not tuned for 
chameleon afterglow searches, one can probe a so far unaccessible region of the parameter space
of chameleon field theories.

\section{Chameleon Cavity \`a la Fabry-P\'erot}\label{sec:CCalaFP}

We have seen, that in the case of a ``chameleon gas'' the afterglow rate of photons is limited to 
$I_\gamma(t=0) < P_{\gamma\rightarrow\phi} I_\gamma$ (cf. Eq.~\eqref{eqn:Idet}). 
We will show in this section that, in principle, 
interference effects of the chameleon wave with the laser can increase the rate of regenerated photons 
and the experimental sensitivity. Ignoring the experimental problem of how to stabilize a chameleon 
wave in a Fabry-P\'erot cavity for more than about $100$ seconds, we will show that coherent interactions 
might increase the afterglow rate up to the laser intensity $I_\gamma$. In the following we will derive 
the amplitude of the right-moving chameleon plane wave taking into account contributions from preceding 
reflections.

Initially, as we switch on the laser beam, we assume the system to be in a pure right-moving photon state 
$\psi_0=(1,0)^T$ at the entrance window at $z=0$ normalized to the laser amplitude. At the far-side 
window at position $z=\ell$ the state has evolved by 
$\mathcal{Q}^T(\varphi)\mathcal{M}'(\ell)\mathcal{Q}(\varphi)\equiv\mathcal{M}_R(\ell)$ 
(see Sect.~\ref{sec:afterglow}). At the window the photon wave escapes and the chameleons are reflected 
back into the cavity. Due to reflection properties and propagation outside the magnetic field 
(see Fig.~\ref{fig:Setup}a) the chameleon field will pick up a phase $\delta_1$ which is in general 
complex\footnote{Sign flips of the chameleon wave are also compensated by this phase.}. The two-level 
state is hence projected after reflection by $R_1=\text{diag}(0,e^{i\delta_1})$ and generates a 
left-propagating chameleon wave. Reaching the near-side mirror this state has further evolved by 
$\mathcal{Q}^T(-\varphi)\mathcal{M}'(\ell)\mathcal{Q}(-\varphi)\equiv\mathcal{M}_L(\ell)$ and is 
projected into a chameleon state by $R_2=\text{diag}(0,e^{i\delta_2})$. In summary, after one cycle 
the system has evolved into a state  
\begin{equation}\label{Cdef}
R_2\mathcal{M}_L(\ell)R_1\mathcal{M}_R(\ell)\psi_0\equiv \mathcal{C}\psi_0 . 
\end{equation}
The matrix $\mathcal{C}$ has the following form to leading order in $\varphi$
\begin{gather*}
\mathcal{C}_{11}=\mathcal{C}_{12}=0\,,\\
\mathcal{C}_{21}=e^{i\delta}\left[-2\sin^2\left(\frac{\ell m^2}{4\omega}\right)\varphi-
i\sin\left(\frac{\ell m^2}{2\omega}\right)\varphi +\mathcal{O}(\varphi^3)\right]\,,\\
\mathcal{C}_{22}=e^{i\delta}\left[1-4\sin^2\left(\frac{\ell m^2}{4\omega}\right)\varphi^2
+2i\left(\frac{\ell m^2}{2\omega}-\sin\left(\frac{\ell m^2}{2\omega}\right)\right)\varphi^2 
+\mathcal{O}(\varphi^4)\right]\,,
\end{gather*}
with a common phase $\delta=\delta_1+\delta_2-2\ell(\omega-\Delta_\phi)$.
The real and imaginary parts of $\mathcal{C}_{22}$ (up to the overall phase) are the familiar contributions 
in the photon-axion system, resulting in a rotation and ellipticity of the laser polarization with respect 
to the magnetic field. Here, however, the role of the axion and photon is interchanged since only the 
former is reflected. From Eqs.~(\ref{prob}) and (\ref{Cdef}) we can derive the relations 
$|\mathcal{C}_{22}|=P_{\phi\to\phi}$ and $|\mathcal{C}_{21}|^2=P_{\phi\to\phi}P_{\gamma\to\phi}$, 
which will become useful in the following.

\begin{figure}[p]
\begin{minipage}[c]{\linewidth}\centering
\includegraphics[width=\linewidth]{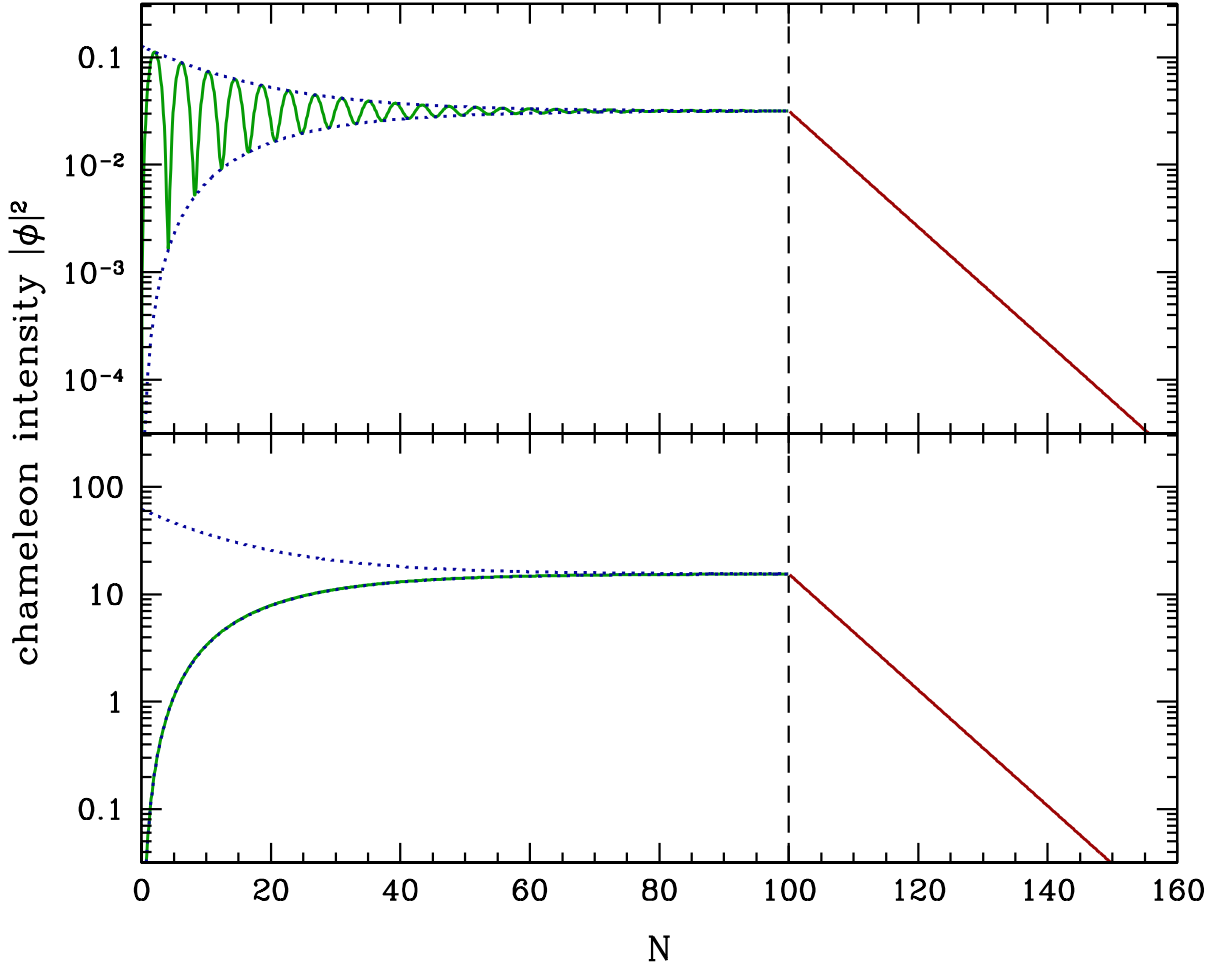}
\end{minipage}
\caption[]{\small Coherent evolution of the chameleon amplitude with number of cycles $N$. For illustration, 
we took $M=100$~GeV, $m=10\,\mu$eV, $\ell=10$~m, $B=5$~T and $\omega=1$~eV. The upper panel shows the case 
$\delta_1+\delta_2=0$ and a rapid oscillation. For the lower panel we chose $\delta=0$, 
such that the chameleon amplitudes add up resonantly. In this case the asymptotic amplitude is 
$|\phi_\infty|\sim1/P_{\gamma\to\phi}$.  Also shown is the envelope 
$|\phi_\infty|^2\cdot(1\pm e^{-t\Gamma_\text{load}})^2$ as a dotted line. After $N=100$ cycles the laser 
is switched off (marked as a dashed line) and the chameleon amplitude decays exponentially. Note that we 
use different ordinates in both panels.}\label{fig:chameleonamp}
\end{figure}

\begin{figure}[p]
\begin{minipage}[c]{\linewidth}\centering
\includegraphics[width=\linewidth]{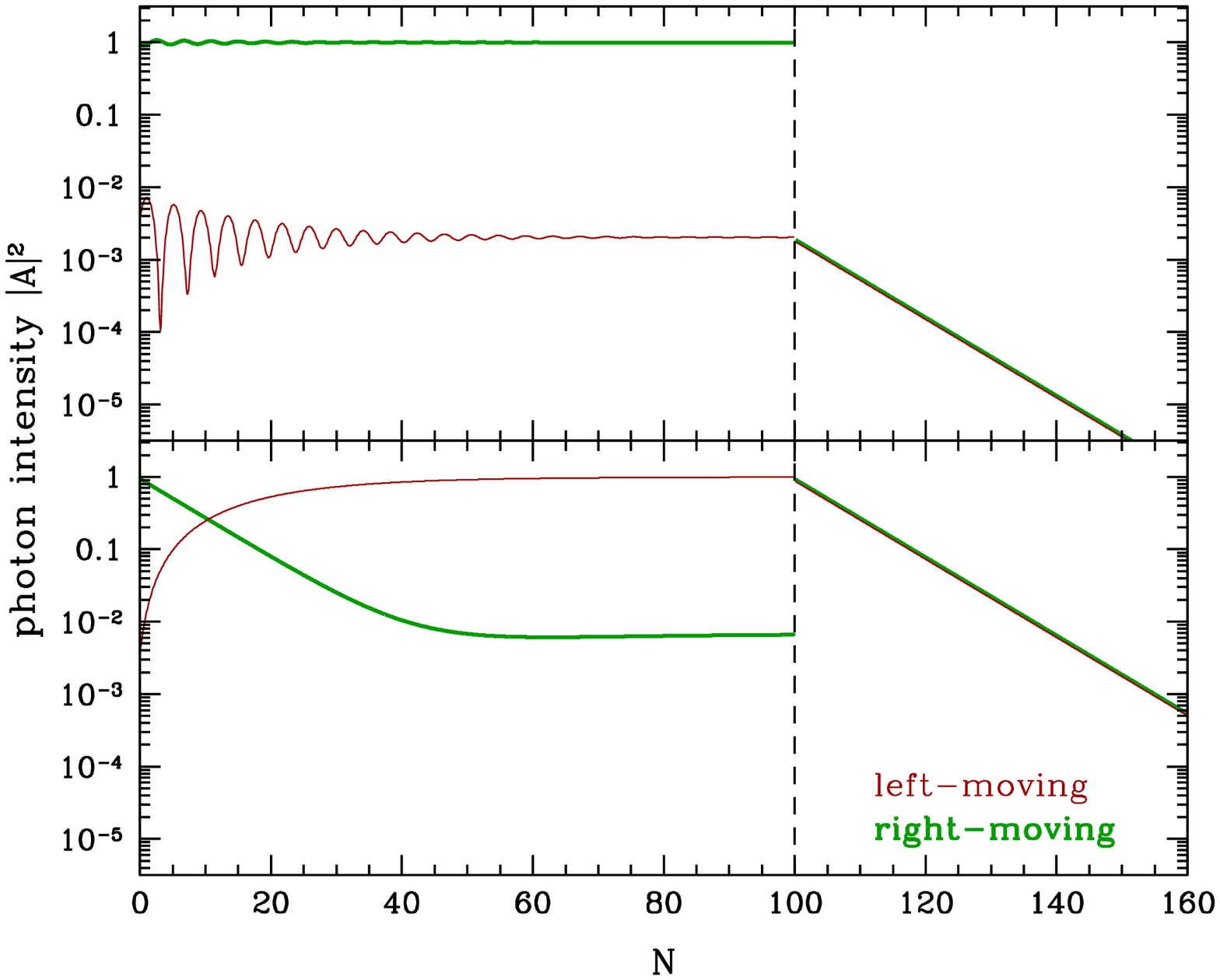}
\end{minipage}
\caption[]{\small Coherent evolution of the photon amplitude with number of cycles $N$ for the benchmark 
points of Fig.~(\ref{fig:chameleonamp}). We show the intensity of the right-moving (left-moving) 
photons transmitted at $z=\ell$ ($z=0$) from the cavity as a thick green (thin red) line. In the resonant 
case (lower panel) the cavity acts as a mirror: 
The amplitude of right-moving photons is suppressed, whereas the left-moving photons reach the laser 
intensity. Again, after $N=100$ cycles the laser is switched off and photons are emitted with exponentially 
decaying intensities to both sides.  In the resonant case the intensity of the beam reaches the 
laser intensity.}\label{fig:photonamp}
\end{figure}

After the first cycle the right-moving two-level system receives a new contribution from the un-propagated 
laser light $(1,0)^T$. Hence, for the second cycle we have to evolve $(\mathcal{C}+\mathbf{1})\psi_0$ with 
$\mathcal{C}$ and so on. After $N$ cycles we arrive at a right-moving state
\begin{equation}
\psi_N=\sum_{k=0}^N\mathcal{C}^k\psi_0= 
\left(1,\mathcal{C}_{21}\frac{1-\mathcal{C}_{22}^N}{1-\mathcal{C}_{22}}\right)^T\qquad 
(\mathcal{C}^0\equiv\mathbf{1})\,.
\end{equation}
The geometric series converges for $N\to\infty$ since $|\mathcal{C}_{22}|<1$ and results in the asymptotic 
state
\begin{equation}
\psi_\infty= \left(1,\frac{\mathcal{C}_{21}}{1-\mathcal{C}_{22}}\right)^T=\left(1,\phi_\infty\right)^T\,.
\end{equation}
The amplitude of the right-moving chameleon wave at $z=0$ after $N$ previous cycles is given as
\begin{equation}\label{Iphi}
|\phi_N|^2 = \left|\mathcal{C}_{21}\frac{1-\mathcal{C}^N_{22}}{1-\mathcal{C}_{22}}\right|^2 
= |\phi_\infty|^2\cdot\left|1-\mathcal{C}^N_{22}\right|^2\,.
 \end{equation}
Since $1-|\mathcal{C}_{22}|^N\leq|1-\mathcal{C}_{22}^N|\leq1+|\mathcal{C}_{22}|^N$
the oscillation $|\mathcal{C}_{22}|^N$ of the squared amplitude around its asymptotic value is damped 
according to $|\mathcal{C}_{22}|^N = e^{-t\Gamma^\text{coh}_\text{load}}$ with
\begin{equation}\label{Tchacoh}
\Gamma_\text{load}^\text{coh} = -\frac{\ln|\mathcal{C}_{22}|}{2\ell}=-\frac{\ln P_{\phi\to\phi}}{2\ell}
=-\frac{\ln\left(1-P_{\gamma\to\phi}\right)}{2\ell} \simeq\frac{P_{\gamma\to\phi}}{2\ell}\,.
\end{equation}
Figure~\ref{fig:chameleonamp} shows the chameleon amplitude for two different set-ups, where we use for 
illustration an (unacceptably) large chameleon coupling corresponding to $M=100$~GeV. 
The configuration in the upper panel 
has a strong oscillation with cycles $N$ caused by the phase of $\mathcal{C}_{22}$. In general, this 
leads to a suppression of the chameleon amplitude at $z=0$. The oscillation is reduced in the resonant case, 
{\it i.e.} $\mathcal{C}_{22}\simeq|\mathcal{C}_{22}|$, which corresponds to  
$\delta\simeq2\pi\mathbb{Z}$ and $\ell\Delta_\phi\varphi^2\ll1$. 
In the case $\ell\Delta_\phi\ll1$, {\it i.e.} 
$\ell\ll\ell_\text{coh}$, we arrive at an asymptotic amplitude of 
\begin{equation}
|\phi_N|^2 \simeq \frac{P_{\phi\to\phi}}{P_{\gamma\to\phi}}
\left(1-P^N_{\phi\to\phi}\right)^2\xrightarrow{N\to\infty} 
\frac{P_{\phi\to\phi}}{P_{\gamma\to\phi}}\simeq \frac{1}{P_{\gamma\to\phi}}\,.
\end{equation}
Such a situation is shown in the lower panel of Fig.~\ref{fig:chameleonamp} with the same experimental and 
model parameters used in the upper panel. 

Once the ``cavity'' is filled with chameleons we switch off the laser field to search for photons regenerated 
in the magnetic field. For a chameleon state $\psi_\phi = (0,\phi)^T$ we have per cycle:
\begin{equation}
\mathcal{C}\psi_\phi = \mathcal{C}_{22}\psi_\phi\,. 
\end{equation}
Per cycle $\Delta t\simeq2\ell$ ($m\ll\omega$), the amplitude of the chameleon beam at $z=0$ decreases as
\begin{equation}
\Delta \dot{N}_\phi = (1-|\mathcal{C}_{22}|^2)\dot{N}_\phi
\end{equation}
This defines an \emph{unloading} rate $\Gamma^\text{coh}_\text{unload}$ as
\begin{equation}\label{Tdiscoh}
\Gamma^\text{coh}_\text{unload} = \frac{1-|\mathcal{C}_{22}|^2}{2\ell}
= \frac{1-P_{\phi\to\phi}^2}{2\ell}\simeq \frac{P_{\gamma\to\phi}}{\ell}\,.
\end{equation}
Note that to leading order $\Gamma^\text{coh}_\text{unload}\simeq2\Gamma^\text{coh}_\text{load}$, in contrast 
to the incoherent case, where there is only one rate for loading and unloading: The loading of the cavity 
takes twice the time as in the case of the gas. 

The photon flux out of the cavity in both configurations is shown in Fig.~\ref{fig:photonamp}. In the upper 
panel, the \emph{off-resonant} case, the photon flow is dominated by the laser light and drops by several 
orders of magnitude after switching off the laser. The situation changes completely in the \emph{resonant} 
case shown in the lower panel. Here, the cavity acts as a mirror: in the stationary limit, the laser light 
is almost completely reflected into left-moving waves. After we have switched off the laser the photons 
regenerated by the chameleons have the intensity of the laser light. 

Note that the larger the final chameleon amplitude ($\propto P^{-1}_{\gamma\to\phi}$) gets the longer it 
takes ($\propto P^{-1}_{\gamma\to\phi}$) to reach it. In a realistic experiment we can only expect a 
limited number of cycles $N\sim\Delta t/2\ell$ for the coherent building up of the amplitude. In this case 
the amplitude will have reached a level
\begin{equation}
|\phi_N|^2 \simeq 
\frac{P_{\phi\to\phi}}{P_{\gamma\to\phi}}\left(1-P^N_{\phi\to\phi}\right)^2\xrightarrow{P_{\gamma\to\phi}\ll1} 
N^2P_{\gamma\to\phi}\simeq\left(\frac{\Delta t}{2\ell}\right)^2P_{\gamma\to\phi}\,.
\end{equation}
The detected afterglow rate of photons at one window is then
\begin{equation}
I_\gamma^{\text{det}}(t=0)  \simeq  
f  \left(\frac{\Delta t}{2\ell}\right)^2 P^2_{\gamma\rightarrow\phi}
 I_\gamma\,,
\label{eqn:Idet2}
\end{equation}
which is about a factor $\Delta t/(4\ell)$  larger than in the case of the chameleon gas 
(see Eq.~(\ref{eqn:Idet})).

So far, we have left aside the experimental problem of a stabilization of the chameleon Fabry-P\'erot 
cavity. 
Considering the lower panel of Fig.~\ref{fig:photonamp} one notices that the optical reflectivity increases 
in the resonant case. This damping of the laser amplitude might be used as a feedback signal for an active 
stabilization in an experiment. However, an actual realization of this concept is beyond the scope of this 
paper.

\section{Conclusions}

In this paper we have proposed a hitherto unnoted signature of chameleon field theories 
--- an afterglow effect --- 
which could be easily exploited in conventional axion-like particle search experiments.

In strong magnetic fields, laser photons can convert to chameleon particles which can be trapped inside a 
vacuum cavity.  We have shown that these particles are expected to form a gas inside the cavity, and that they
might reveal themselves as an afterglow after the laser is switched off. 
In general, for long enough loading times, a stationary state is achieved, where the chameleon production 
in the laser beam and the loss via back-conversion to photons is balanced. In this case, the afterglow rate 
only depends on the initial production rate and not on the details of the experiment. 
This kind of experiment is in principle sensitive to couplings below
$1/M \sim 10^{-8}\,\text{GeV}^{-1}$, which 
is orders of magnitude smaller than the sensitivity of typical laser polarization experiments.
Its sensitivity also compares favorably with Casimir force experiments.  
The main constraints on afterglow experiments come from the required loading times 
which can easily become orders of magnitudes larger than $\sim\mathcal{O}(100\,\text{s})$.  

We have also sketched the concept of a Fabry-P\'erot cavity for the chameleons. With an    
appropriate tuning of phase shifts the sensitivity of the experiment compared to the case of a chameleon 
gas might be further increased by a factor $\Delta t/4\ell$ in the case where the loading time $\Delta t$ 
is much smaller than the inverse loading rate $\Gamma$. In the (academic) case $\Delta t\Gamma\gg1$ the 
cavity will act as a mirror for the incoming photon beam and the rate of afterglow photons in the unloading 
phase of the cavity reaches the intensity of the initial laser beam.
However, in an experimental realization of this concept the time scale  
$\Delta t$ will be limited by the effective finesse of the chameleon 
cavity, {\it i.e.}~the maximal number of cycles achievable.

Open questions concern the possible thermalization of the chameleon gas, and the question under which 
circumstances the particle picture breaks down due to strong coupling effects. The answers will most 
probably depend strongly on the specific potential $V(\phi)$ and might lead to strong constraints on 
chameleon theories. We will consider this elsewhere~\cite{We:2007}.


\bibliographystyle{h-physrev3}
\frenchspacing
\bibliography{refs}

\begin{thebibliography}{10}

\bibitem{Wetterich:1987fm}
C.~Wetterich,
\newblock Nucl. Phys. {\bf B302}, 668 (1988).

\bibitem{Ratra:1987rm}
B.~Ratra and P.~J.~E. Peebles,
\newblock Phys. Rev. {\bf D37}, 3406 (1988).

\bibitem{Caldwell:1997ii}
R.~R. Caldwell, R.~Dave, and P.~J. Steinhardt,
\newblock Phys. Rev. Lett. {\bf 80}, 1582 (1998), astro-ph/9708069.

\bibitem{Khoury:2003aq}
J.~Khoury and A.~Weltman,
\newblock Phys. Rev. Lett. {\bf 93}, 171104 (2004), astro-ph/0309300.

\bibitem{Khoury:2003rn}
J.~Khoury and A.~Weltman,
\newblock Phys. Rev. {\bf D69}, 044026 (2004), astro-ph/0309411.

\bibitem{Brax:2004qh}
P.~Brax, C.~van~de Bruck, A.-C. Davis, J.~Khoury, and A.~Weltman,
\newblock Phys. Rev. {\bf D70}, 123518 (2004), astro-ph/0408415.

\bibitem{Mota:2006ed}
D.~F. Mota and D.~J. Shaw,
\newblock Phys. Rev. Lett. {\bf 97}, 151102 (2006), hep-ph/0606204.

\bibitem{Mota:2006fz}
D.~F. Mota and D.~J. Shaw,
\newblock Phys. Rev. {\bf D75}, 063501 (2007), hep-ph/0608078.

\bibitem{Brax:2007vm}
P.~Brax, C.~van~de Bruck, A.-C. Davis, D.~F. Mota, and D.~Shaw,
\newblock arXiv:0709.2075 [hep-ph].

\bibitem{Jaeckel:2006xm}
J.~Jaeckel, E.~Masso, J.~Redondo, A.~Ringwald, and F.~Takahashi,
\newblock Phys. Rev. {\bf D75}, 013004 (2007), hep-ph/0610203.

\bibitem{Brax:2007ak}
P.~Brax, C.~van~de Bruck, and A.-C. Davis,
\newblock hep-ph/0703243.

\bibitem{Brax:2007hi}
P.~Brax, C.~van~de Bruck, A.-C. Davis, D.~F. Mota, and D.~J. Shaw,
\newblock arXiv:0707.2801 [hep-ph].

\bibitem{Cameron:1993mr}
[BFRT Collaboration], R.~Cameron {\em et~al.},
\newblock Phys. Rev. {\bf D47}, 3707 (1993).

\bibitem{Rizzo:2006bm}
[BMV Collaboration], C.~Rizzo,
\newblock 2nd ILIAS-CERN-CAST Axion Academic Training 2006,
  http://cast.mppmu.mpg.de/.

\bibitem{Zavattini:2005tm}
[PVLAS Collaboration], E.~Zavattini {\em et~al.},
\newblock Phys. Rev. Lett. {\bf 96}, 110406 (2006), hep-ex/0507107.

\bibitem{Zavattini:2007ee}
[PVLAS Collaboration], E.~Zavattini {\em et~al.},
\newblock arXiv:0706.3419 [hep-ex].

\bibitem{Pugnat:2006ba}
[OSQAR Collaboration], P.~Pugnat {\em et~al.},
\newblock CERN-SPSC-2006-035.

\bibitem{Chen:2006cd}
[Q\&A Collaboration], S.-J. Chen, H.-H. Mei, and W.-T. Ni,
\newblock hep-ex/0611050.

\bibitem{Ehret:2007cm}
[ALPS Collaboration], K.~Ehret {\em et~al.},
\newblock hep-ex/0702023.

\bibitem{ALPS:2007}
[ALPS Collaboration],
\newblock http://alps.desy.de/.

\bibitem{GammeV:2007}
[GammeV Collaboration],
\newblock http://gammev.fnal.gov/.

\bibitem{Baker:2006li}
[LIPSS Collaboration], K.~Baker,
\newblock 2nd ILIAS-CERN-CAST Axion Academic Training 2006,
  http://cast.mppmu.mpg.de/.

\bibitem{Cantatore:2006pv}
[PVLAS Collaboration], G.~Cantatore,
\newblock 2nd ILIAS-CERN-CAST Axion Academic Training 2006,
  http://cast.mppmu.mpg.de/.

\bibitem{Brax:2007}
H.~Gies, D.~F. Mota, and D.~J. Shaw,
\newblock arXiv:0710.1556 [hep-ph].

\bibitem{Raffelt:1987im}
G.~Raffelt and L.~Stodolsky,
\newblock Phys. Rev. {\bf D37}, 1237 (1988).

\bibitem{We:2007}
A.~Ringwald, L.~Schrempp, and C.~Weniger,
\newblock work in progress.

\end{thebibliography}
 
 
\end{document}